\documentclass[twocolumn,,amsmath,amssymb, aps, showpacs]{revtex4}

\usepackage{graphicx}
\usepackage{dcolumn}
\usepackage{bm}
\newcommand{\etal}{\emph{et al.}}
\newcommand{\be}{\begin{equation}}
\newcommand{\ee}{\end{equation}}
\newcommand{\bfig}{\begin{figure}}
\newcommand{\efig}{\end{figure}}
\newcommand{\incl}{\includegraphics}

\begin{document}

\title{Bulk band gap and surface state conduction observed in voltage-tuned crystals of the topological insulator Bi$_2$Se$_3$}

\author{J. G. Checkelsky$^{1,\dagger}$, Y. S. Hor$^{2,\ddagger}$, R. J. Cava$^2$ and N. P. Ong$^1$
} 
\affiliation{Department of Physics$^1$ and Department of Chemistry$^2$,\\ 
Princeton University, New Jersey 08544, U.S.A.}

\date{\today}


\begin{abstract}
We report a transport study of exfoliated few monolayer crystals of topological insulator Bi$_2$Se$_3$ in an electric field effect (EFE) geometry.  By doping
the bulk crystals with Ca, we are able to fabricate devices with sufficiently low bulk carrier density to change the sign of the Hall density with 
the gate voltage $V_g$.  We find that the temperature $T$ and magnetic field dependent transport properties in the vicinity of this $V_g$ can be explained by a bulk channel with 
activation gap of approximately 50 meV and a relatively high mobility metallic channel that dominates at low $T$.  The conductance (approximately 2 $\times$ 7$e^2/h$), weak anti-localization, and metallic resistance-temperature profile of the latter lead us to identify it with the protected surface state.  The relative smallness of the observed gap implies limitations for EFE topological insulator
devices at room temperature.
\end{abstract}

\maketitle                   
Topological insulators (TIs) are a new phase of matter which is electrically insulating in the bulk but has unusual conducting surface states (SS) \cite{Fu07, Moore07, Bernevig06, FuKane07, QiHughes08}.  The SS are spin polarized, protected from scattering by non-magnetic impurities, and have an approximately linear energy-momentum dispersion at low energy.  These properties make them relevant for applications ranging from improved spintronic devices to producing analogs of exotic high energy particles potentially useful for quantum computing \cite{HasanReview}.  Particular interest has been focused on 3D TIs based on Bi, where the topologically non-trivial surface 2D electron gas has been positively identified by angle-resolved photoemission spectroscopy (ARPES) in Bi-Sb \cite{Hsieh08}, Bi$_2$Se$_3$ \cite{Xia09}, and Bi$_2$Te$_3$ \cite{Shen09}.  Proposals for realizing devices that harness the aforementioned properties of the SS most often involve electrical transport \cite{FuKane09, FuKane09b, Akhmerov09, Franz09, Linder10}.  Thus far, characterizing or even identifying the SS in such experiments has proven difficult.  Because transport is a bulk sensitive measurement, even a small conductivity from imperfections in the bulk overwhelms the surface contribution because of the geometric advantage of the former.  The materials obstactles to placing the chemical potential $\mu$ in the bulk band gap of the TI are at present the most serious obstacle to realizing TI devices.  

\bfig[t]            
\incl[width=8cm]{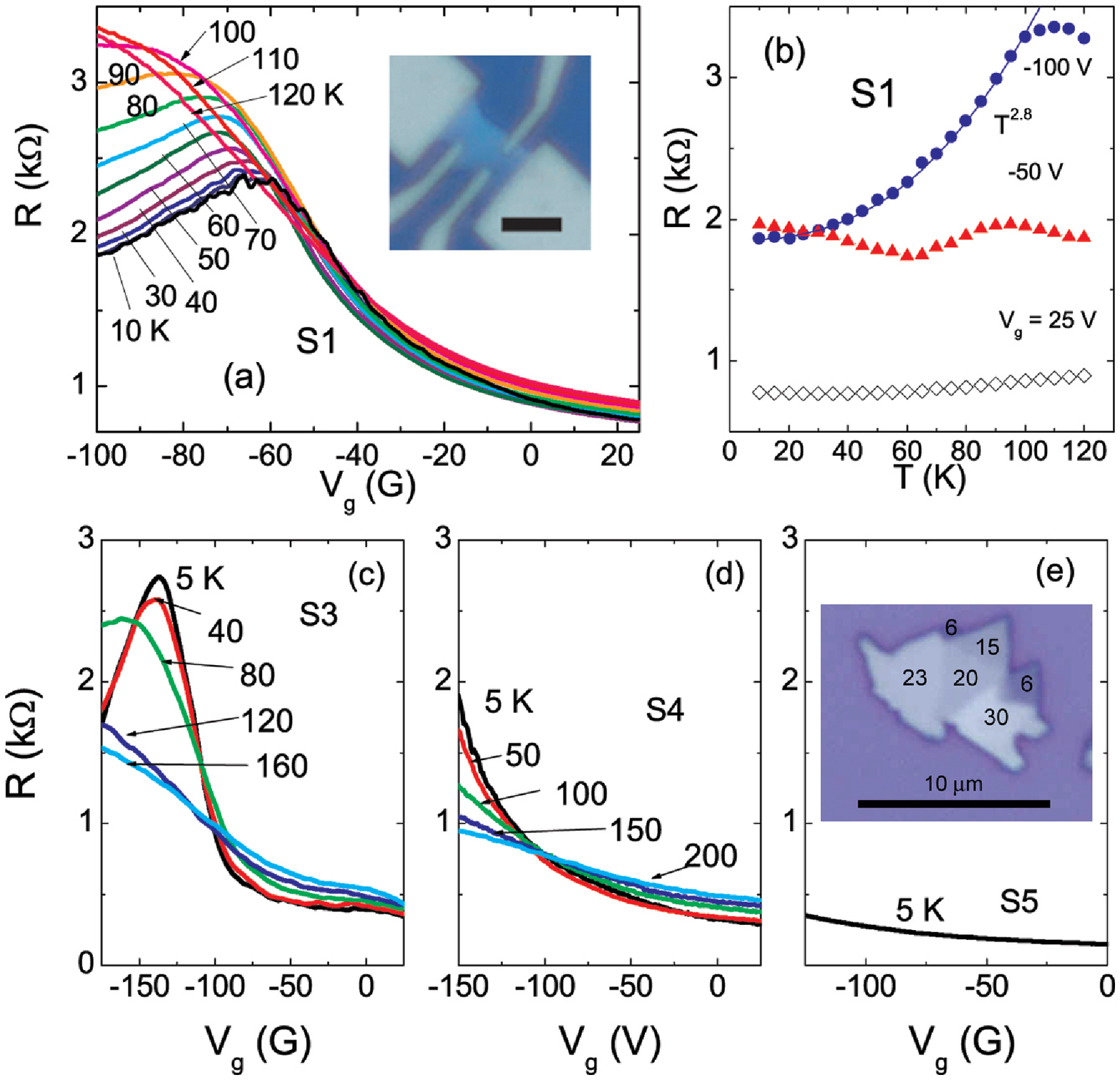} 
\caption{  (a) $R$ vs. $V_g$ at selected $T$ between
10 and 120 K for sample S1.  The inset is a photo of device S1 with scale-bar (4 $\mu$m).  (b) Cuts
of $R$ vs. $T$ with $V_g$ fixed at -100, -50 and +25 V.  Gating results for samples S3, S4, and S5
are shown in panels (c), (d), and (e).  An optical image of a cleaved crystal with various step heights with thickness labels measured by atomic
force microscopy (in nm) is shown in the inset of (e).  
\label{figR} 
}
\efig

Recent measurements in high magnetic fields have identified the surface state as a parallel conductance channel with the bulk \cite{oblig1, oblig2, oblig3}.  
Several experiments have been recently undertaken to circumvent the bulk contribution to transport in TIs by preparing Bi$_2$Se$_3$ in thin film form \cite{Chen10, MIT, Kong10, Minhao10}.  However, in both bulk crystals and nanocrystals of Bi$_2$Se$_3$, $\mu$ is usually pinned to the conduction band (CB) due to the presence of Se vacancies \cite{Hyde74, MIT}.  This large remnant electron density is too large to remove by the electrostatic gating techniques available for thin films.  Thus, while thin film experiments have been able to show signs of surface phenomena, namely remnants of weak anti-localization \cite {Chen10, Minhao10}, basic transport properties of TIs have remained elusive as the bulk carriers cannot be effectively removed.   

Here we report experiments on Ca passivated crystals of Bi$_2$Se$_3$ in a field effect transistor device that allow us to successfully suppress the bulk states and investigate the transport properties of the SS.  We mechanically exfoliated crystals of Ca doped Bi$_{2}$Se$_{3}$ (up to 0.5\% substituted for Bi) onto doped silicon wafers coated with 300 nm SiO$_{2}$.  The Ca doping results in hole doped bulk crystals \cite{Hor09, Check09}, which after fabrication of contacts via e-beam lithography (Cr/Au) in the Hall geometry have $\mu$ relatively close to the bulk band gap on the CB side.  In samples with uniform thickness $d$, a negative $V_g$ larger than -150 V can be applied without triggering breakdown.  For $d<$ 10 nm, from electrostatic arguments this suffices to move $\mu$ deep into the gap for bulk carrier densities 10$^{18}$ - 10$^{19}$ cm$^{-3}$.

The use of 300 nm SiO$_{2}$ allows optical identification of thin crystals.  It was found that for crystals thinner than $\sim$ 30 nm the observed color of the cleaved crystals darkens toward the color of the substrate.  This is similar to what has been found in graphene and other 2D crystals \cite{Geim04}.  Cross correlating the optical images with atomic force microscope (AFM) images, we were able to generate a mapping between color and thickness useful for identifying thin crystals optically within an approximate thickness $\pm$ 2 nm.  An optical image of a crystal with varying thickness measured by both AFM and optical microscopy is shown in the inset of Fig \ref{figR}e.  We study crystals down to 5 nm; ARPES experiments on thin films of Bi$_2$Se$_3$ have shown the SS persist to this thickness \cite{arpesThin}.

Table \ref{Table1} shows parameters for five different devices.  Devices of thickness $t \leq 20$ nm exhibit similar characteristics.  For these devices an applied gate voltage can change the Hall response from electron-like to hole-like at the neutral voltage $V_{N}$.  Each exhibits a maximum at low $T$ in the resistance per square $R_{\textrm{max}}$ of $2-3 $ k$\Omega$, as shown for samples S1 and S3 in Fig. \ref{figR}a and \ref{figR}c, respectively.  For thicker crystals such as devices S4 and S5, we reach the threshold for breakdown of the SiO$_{2}$ gate dielectric (typically -180 V to -200 V) without observing such a peak (see Fig. \ref{figR}d and \ref{figR}e).  The progressively thicker crystals show systematically less response to the applied gate voltage.  It is found that the act of cleaving and device fabrication reduces the carrier mobility and increases the electron density.  Device S5 (see Fig. 1e and Table I) has a 0.3 K mobility approximately 1090 cm$^2$ V$^{-1}$s$^{-1}$ with 6 $\times$ 10$^{18}$ e$^{-}$ cm$^{-3}$ was cleaved from a bulk crystal measured to have a mobility 7000 cm$^2$ V$^{-1}$s$^{-1}$ with 2 $\times$ 10$^{18}$ e$^{-}$ cm$^{-3}$.  It is unclear if the cleaving process or lithographic process is responsible for this degradation.  From here we focus on the results of samples S1 and S2.  

\begin{table}[htb]
\begin{center}
\begin{tabular}{|c||cccc|}
	\hline
		  &  $t$   &  $V_{N}$  & $\mu_b$   & $R_{\textrm{max}}$ \\
units	& nm	 &	V 			 &	cm$^{2}/$Vs	&	k$\Omega$		 \\
	\hline \hline
S1   & 10 $\pm$ 2 & -80   & 330 & 2.4\\
S2   & 6  $\pm$ 2 & -90   & 830 & 2.3\\
S3   & 20 $\pm$ 4 & -170  & 720 & 2.8\\
S4   & 30 $\pm$ 5 & $<$ -175 & 910 & 1.9\\
S5   & $>40$ & $<$ -175 & 1090 & 0.3\\
	\hline
\end{tabular}
\caption{\label{Table1} Sample parameters.  $t$ is the sample thickness estimated from an optical image of the cleaved crystal (see inset of Fig \ref{figR}e).  $V_{N}$ is the gate voltage required to reach zero Hall voltage at 5 K.  $\mu_b$ is the Hall mobility in the conduction band, calculated at zero gate voltage.  $R_{\textrm{max}}$ is the maximum resistance recorded over the measured gating range.}
\end{center}
\end{table}

Figure \ref{figR}a displays traces of the zero-field resistance 
$R\equiv R_{xx}$ vs. $V_g$ at selected $T$ from 10 to 120 K (sample S1).  
For $V_g>$ -55 V, $R$ is nearly $T$ independent, characteristic of a bad metal
in which disorder scattering is dominant (Fig. \ref{figR}b). Surprisingly, if $\mu$ is tuned lower ($V_g<$ -55 V), $R$ displays a power-law variation $R\sim T^{\alpha}$ ($\alpha$=2.8) that extends to 110 K, where $R$ displays a broad maximum.  
We show below (Eq. \ref{eq:G}) that both the gate-induced
transformation (from a $T$-independent $R$ to one with power-law
variation) and the existence of the broad peak in $R$ are accounted for by a 
2-band model comprised of a bulk band in parallel with a surface channel
with higher mobility at this large negative gate voltage.  When $V_g$ is near -55 V, $\mu$ appears to be at the CB band edge.

\bfig[t]            
\incl[width=8cm]{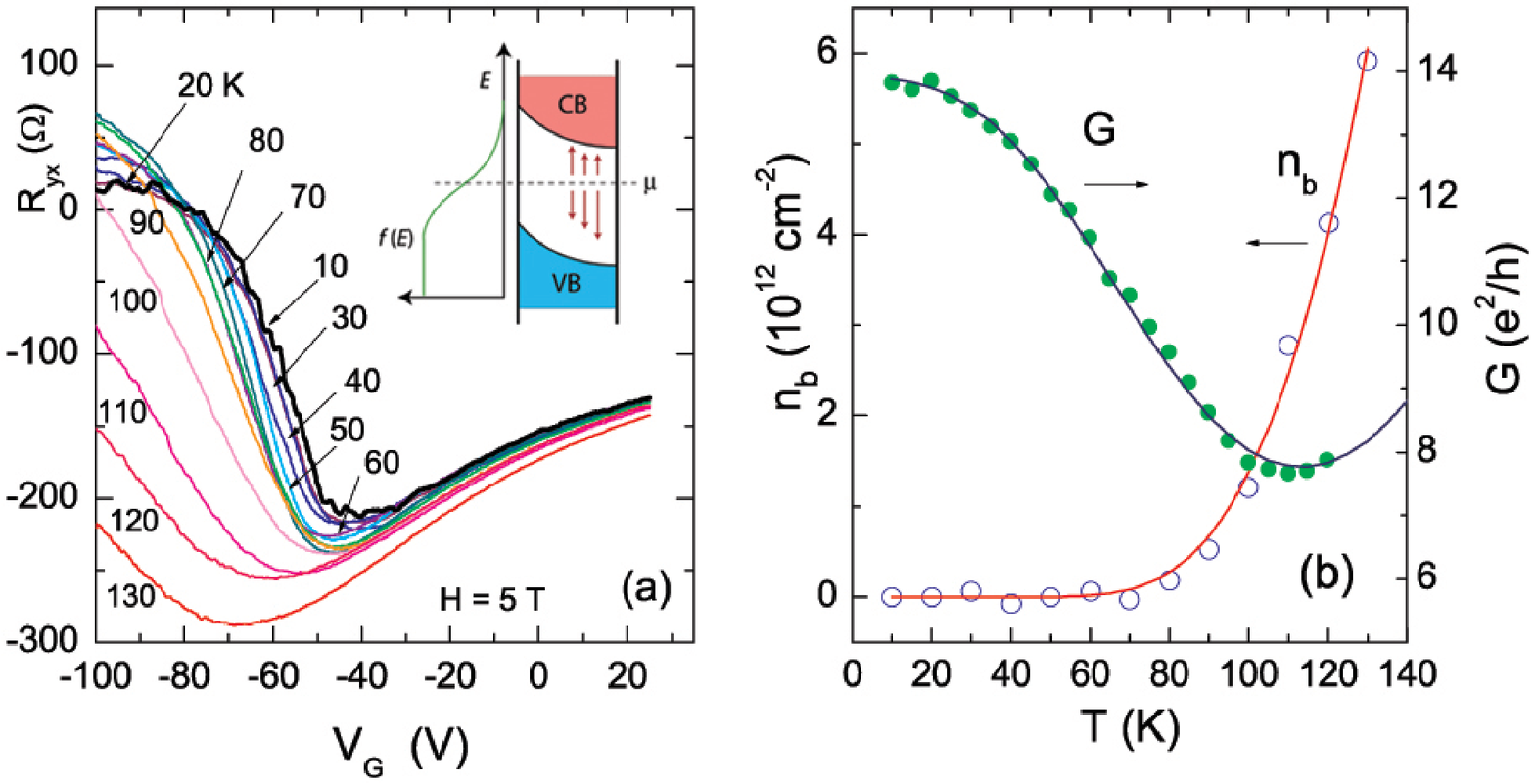} 
\caption{
(a) Hall resistance $R_{yx}$ vs. $V_g$ at 
selected $T$ from 10 K (bold curve) to 130 K for sample S1. The sketch (inset) shows the
strong band-bending induced by $V_g<$-80 V.  The thermal excitation of 
carriers is indicated by red arrows ($f(E)$ is the Fermi-Dirac distribution). (b) Fit of the conductance 
$G$ at $V_g$=-100 V (solid circles) to Eq. \ref{eq:G},
with $G^s(0)$ = 13.9 $e^2/h$.  Values of
$n_b(T)$ (open circles) inferred from 
$R_{yx}$ and $\Delta V_g$ are plotted as open circles, along with the fit to the 
activated form $n_b = n_{b0}\rm{e}^{-\Delta/T}$ ($\Delta$ = 640 K).
\label{figHall} 
}
\efig

The Hall resistance $R_{yx}$ confirms that the carriers are $n$-type
 for $V_g >$  -55 V (Fig. \ref{figHall}a).  $R_{yx}$ 
is $T$ independent, consistent with a bulk metal.  Additionally, $R_{yx}(V_g)$ is linear
as we would expect from a simple electron band being emptied by the gate (induced charge $n_{ind}
 = CV_g/e$, where $C$ the capacitance per unit area of the gate dielectric).
As $V_g$ is lowered below -55 V, the curves of $R_{yx}$ vs. $V_g$ 
display a strong dependence on $T$.  Below 30 K, 
$R_{yx}$ changes sign near $V_g$ = -80 V.  We can understand this in terms of $\mu$ moving below the SS Dirac point or near the VB edge.
It is important to note that the gating does not rigidly shift $\mu$ in the device.
For samples with $d\ll D$ (the depletion length), the bands 
are strongly bent upwards when $V_g$ is large and negative 
(inset in Fig. \ref{figHall}a).  
As $\mu$ is grounded at the drain, 
it falls within the gap inside the crystal.  
While the specific shape of the bending is difficult to calculate, we can adopt the most simple limit
by treating the system as a metal - (doped) semiconductor interface in the full depletion approximation.  The Poisson equation for the potential $\phi$, charge density $n$ (estimated as 10$^{19}$ e$^{-}$ cm$^{-3}$ from $R_{yx}$), and dielectric constant $\epsilon$ as a function of the distance from the interface $z$ is then $\phi(z) = (en/2\epsilon) (D^{2} - (D-z)^{2}))$.  The curvature of the band is that depicted in Fig. \ref{figHall}a for a 10 nm crystal that should be preserved in more realistic calculations, which will be important for analyzing the role of thermally activated carriers.

We assume that, when $\mu$ is inside the gap, 
the observed conductance $G=1/R$ is the parallel combination
\be
G(T) = G^s(T) + G^b(T),
\label{eq:G} 
\ee
where $G^s(T) = n_s e \mu_s$ is the surface conductance, with $n_s$
the two-dimensional (2D) density and $\mu_s\sim T^{-\alpha}$ the mobility
($e$ is the electron charge and $\alpha$ = 2.8).  To account for disorder
scattering, we write $G^s(T) = G^s(0)/[1+ AT^{\alpha}]$, and treat 
the zero-$T$ surface conductance $G^s(0)$
and $A$ as adjustable parameters.
The second term $G^b(T) = n_b e \mu_b$ is a bulk conductance resulting from 
thermal excitation of carriers into the CB described by the 
density $n_b(T) = n_{b0}{\rm e}^{-\Delta/k_BT}$ (arrows in inset, Fig. \ref{figHall}a).  The mobility $\mu_b$ at the bottom of the CB
is taken to be a constant (disorder-scattering dominant).
We note that, in addition to $\mu$ originally being in the CB, the curvature of the band 
bending in the case of a remnant electron pocket favors excitation in to the CB for devices of finite $d$.  Integrating the
excited carriers $n_{exc} \sim \int^{d}_{0} e (\textrm{e}^{-\Delta_{VB} / k_{B}T} - \textrm{e}^{-\Delta_{CB} / k_{B}T}) dz$ ($\Delta_{CB,VB}$ is the energy difference between the VB and CB band edges and $\mu$) , we can expect the electron-like carriers to dominate due to the direction of the band bending.  

Equation \ref{eq:G} gives a good fit to $G$ vs. $T$
(Fig. \ref{figHall}b).  From the fit, we obtain 
$G^s(0)$ = 13.9 $e^2/h$ ($A$ = 2.5 $\times$ 10$^{-6}$).  This implies a conductance of approximately 7$e^2/h$ per surface.  As $T$ increases from 10 K, inelastic scattering
strongly reduces $G^s(T)$.  Meanwhile, excitation across 
the gap $\Delta$ = 530$\pm$30 K leads to a rapid growth in the bulk $G^b$.
This reverses the decreasing trend of $G(T)$ at 110 K, as shown in Fig. \ref{figR}b.
At 130 K, the 2 terms, $G^b(T)$ = 3.65 $e^2/h$ and $G^s(T)$ = 4.5 $e^2/h$, 
become roughly comparable.  Thus, we show that despite the large intrinsic gap in Bi$_{2}$Se$_{3}$ (300 meV), we must go below 100 K for the device to be 
dominated by surface conduction.  One likely origin of this reduction in energy gap is the band bending
depicted in Fig. \ref{figHall}a, which effictively narrows the gap upon increased gating.  This implies limitations for accessing the SS at room temperature.  

To measure $\Delta$ more accurately, we turn to the 
Hall results in Fig. \ref{figHall}a.  We may obtain 
$n_b(T)$ by studying how $R_{yx}$ depends on both $V_g$ and $T$ 
when $\mu$ is in the gap.  
At the lowest $T$ (bold curve at 10 K), $R_{yx}$ is at the zero-crossing
when $V_g$ = -80 V.  
As $T$ is raised with $V_g$ fixed, the balance condition is unaffected until $T>$80 K, 
whereupon $R_{yx}$ swings to negative values, with the thermal activation 
of carriers into the CB.  However, the balance point can be restored if $V_g$
is made more negative to draw more holes into the crystal (e.g., $V_g\to$ -98 V at 100 K).  From
the gate-shift $\Delta V_g$ needed at each $T$, we calculate the activated density as 
$n_b(T) = \Delta V_gC/e$.  A fit to $n_b(T)$ yields a gap of 640 K (Fig. \ref{figHall}b).  This excitation supports the band bending proposed in Fig. 2a.
Combining the value $n_b\sim 6\times 10^{12}$ cm$^{-2}$ 
at 130 K with the value of $G^b(T)$ inferred from Eq. \ref{eq:G}, we obtain the 
mobility $\mu_b\sim$ 147 cm$^2$/Vs (at the bottom of the CB).

Hence, the gate-driven transformation to a conductor with strong $T$ dependence in both $R$ and $R_{yx}$ 
is explained quantitatively as the opening of a small gap $\Delta\sim$50 meV between $\mu$ and the CB.
Depletion of the CB carriers exposes a high-mobility channel $G^s$.
The possibility that this channel is a bulk impurity band can be excluded because
electrons therein generally have even lower 
mobilities (1-10 cm$^2$/Vs) than obtained for $\mu_b$.

\bfig[t]            
\incl[width=7.5cm]{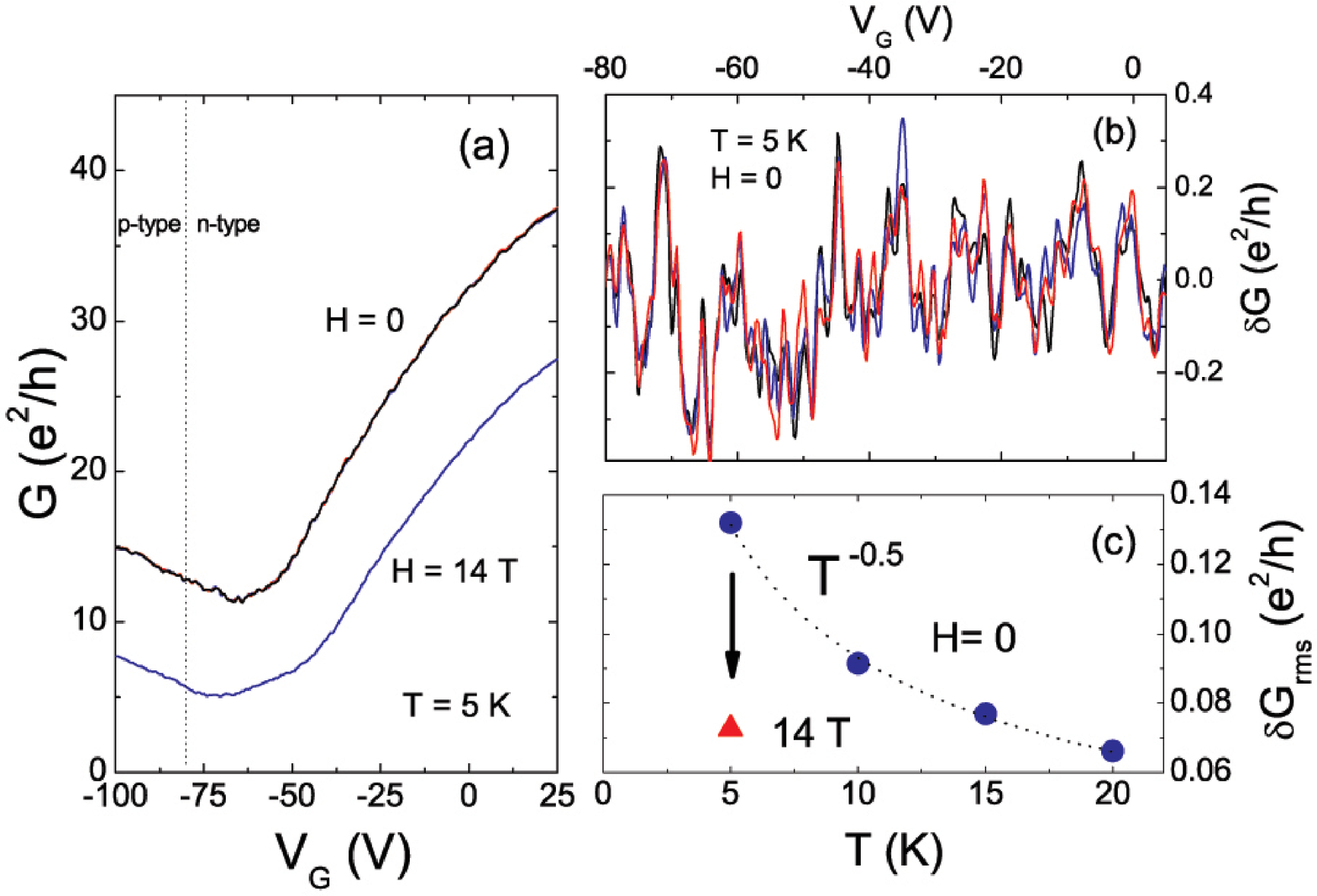} 
\caption{ (a) Conductance $G(T,0)$ in zero $H$ (upper trace) with
$G(T,H)$ at 14 T (lower) at $T$ = 5 K for sample S1. 
In the $H=0$ trace (superposition of 3 traces), retraceable fluctuations are
resolved.
(b) Amplitude of the conductance fluctuation 
$\delta G = G - \langle G\rangle$ (with $\langle G\rangle$ a smooth background)
(c) The $T$ dependence of the rms value $\delta G_{rms}$ (solid circles in inset) 
fits well to $T^{-0.5}$ (dashed curve).  
\label{figG}}
\efig

The enhanced mobility for $G^s$ is confirmed by the transverse magnetoresistance (MR).  
Fig \ref{figG}a compares the zero-field conductance 
$G(T,0)$ at 5 K with the conductance $G(T,H)$ measured in with $\bf H||c$ at 14 T as a function of $V_g$ 
(we suppress $T$ in $G(T,H)$ hereon). The semi-classical MR expression $G(H) = G(0)/[1+(\mu_{ave} H)^2]$ 
may be used to determine the average mobility $\mu_{ave}$.  
At $V_g\sim$20 V ($\mu$ in the CB), the reduction in $G$ by 25$\%$ gives $\mu_{ave}\sim$380 cm$^2$/Vs, 
consistent with the Hall analysis (see Table I). The larger reduction ($60\%$) in the gap region ($V_g\sim$-80 V) 
gives $\mu_{ave}$ = 850 cm$^2$/Vs, which we identify with $\mu_s$.  From $G^{s}$, we can estimate $n_{s} \sim 4, 10$ 
$\times$ 10$^{12}$ cm$^{-2}$ at $V_{g} = -80, 0$ V.

\bfig[t]            
\incl[width=8cm]{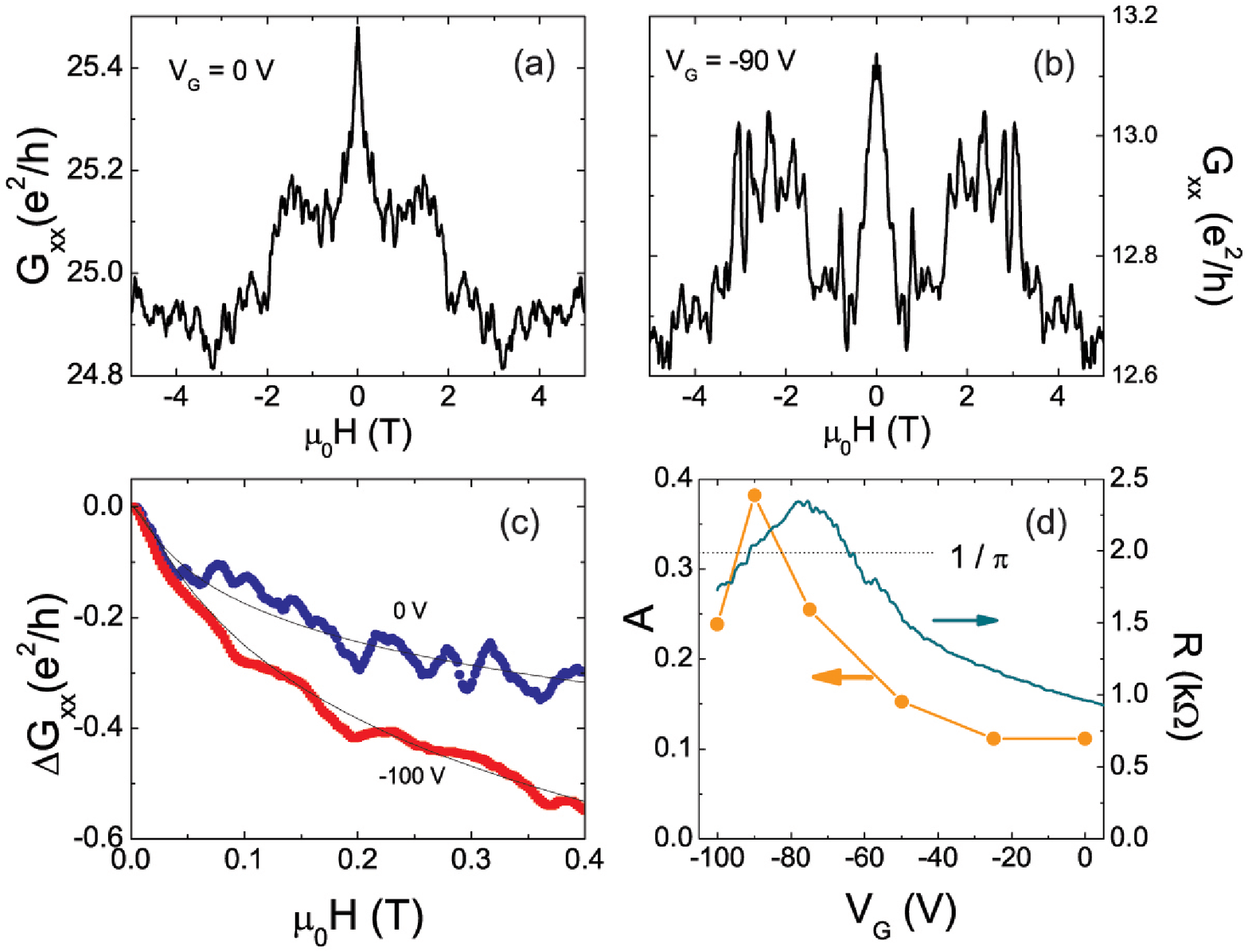} 
\caption{Panels (a) and (b) show conductance fluctuations for sample S2 in $G$ vs. $H$ 
at 0.3 K with $V_g$ fixed at 0 V  and at -90 V. (c) Low-$H$ magnetoconductance 
at $V_g$ = 0 V and $V_g$ =-100 V.  The solid lines are fits to Eq. \ref{eq:anti}. 
(d) Fit parameter $A$ shows a sharp maximum near the 
charge-neutral point.  Dashed line shows the value $A$ = 1/$\pi$
predicted for dominant spin-orbit coupling (see text).  $R(V_{g})$ at 0.3 K is also shown.
\label{fig4} 
}
\efig

At low $T$, fluctuations in $G$ are observed if $H$ or $V_g$ is swept.  
Fig. \ref{figG}b displays the trace of
fluctuating component $\delta G = G - \langle G\rangle$ vs. $V_g$, where
$\langle G\rangle$ is a smooth background.  The amplitude of
$\delta G$ is nominally uniform over the entire range $-100 <V_g< 20$ V.  The amplitude (Fig. \ref{figG}c) follows a $T^{-1/2}$ scaling consistent with a 2D electron system with a dephasing length smaller than the system size \cite{Lee87}.
Recent observation of reproducible fine structure driven by $H$ in both transport
and STM measurements have been reported \cite{Check09, Hanaguri10}.  Here, for $V_G$ = 0 (Fig. \ref{fig4}a), we observe in Sample S2 ($R(V_{g}$) shown in Fig. \ref{fig4}d)
a sharp, narrow peak at $H$ = 0 flanked by mesa-like structures in addition to the fluctuations.  However, the large amplitude fluctuations (rms amplitude up to 6$e^{2}/h$) that we found in bulk samples of the same material previously \cite{Check09} do not survive in these exfoliated devices.  The reason for this and the origin of the detailed structures observed (including the suppression by $H$ shown in Fig. \ref{figG}c) deserve further study.

The sharp anomaly in $G$ at $H$ = 0 in Figs. \ref{fig4}a and \ref{fig4}b is consistent with the quantum
correction arising from anti-localization
given by~\cite{Hikami}
\be
\Delta G_{\textrm{xx}}(H) = A \frac{e^{2}}{h}\left[\ln\frac{H_{0}}{H} - \psi\left(\frac{1}{2} + \frac{H_{0}}{H}\right)\right]
\label{eq:anti}  
\ee
where $H_{0}$ is the dephasing magnetic field, $\psi$ is the digamma function,
and the parameter $A$ is positive for antilocalization.  If spin-orbit coupling is
strong, theory predicts $A=1/2\pi$ \cite{Hikami}, or $2A = 1/\pi$ if we 
have 2 surfaces, as here.  Confining the fit to fields below 0.4 T, we show 2 
traces in Fig. \ref{fig4}c at $V_g$ = 0 and -100 V.  Fits across the full range 
of $V_g$ are shown in Figure \ref{fig4}d.  At $V_N$, $A$ peaks at 0.38, 
which is within 20\% of the expected value $1/\pi$.  Reports of similar experiments
on epitaxially grown films report approximately half of this value, consistent with one
surface state \cite{Chen10, Minhao10}.  We hypothesize the disparity may arise from the different nature
of the strained interface of the epitaxial films and the exfoliated crystal here.  
We also find a peak in the dephasing field $H_{0}$ close to $V_N$ indicating 
a reduction in the dephasing length $\ell_{\phi}$ from 220 nm at zero $V_{g}$ to 67 nm at -90 V.  The value at zero gate is consistent with
those reported in epitaxial films of similar thickness \cite{Minhao10}, while the reduction in $\ell_{\phi}$ has been suggested as evidence for reduced screening and increased electron-electron interaction effects in the low density regime \cite{Chen10}.  The observation of $A \sim 1/\pi$ along with the sign change in $R_{yx}$, supports our main conclusion that, at $V_N$, $\mu$ is well inside the bulk band gap. Notably, even with $\mu$ in the CB, a signature of weak anti-localization persists. 

Electrostatic control of $\mu$ provides a powerful means for comparing the
bulk and surface conductances in chemically doped TIs.  We anticipate that replacing the Au electrodes 
with magnetic or superconducting materials will enable us to probe exotic 
states that have been proposed \cite{FuKane09, FuKane09b, Akhmerov09, Linder10, Franz09}. The current devices allow 
access to surface dominated transport for these experiments below 100 K.

We acknowledge support from the U.S. National Science Foundation (NSF DMR 0819860).
$^\dagger$ Present address: Advanced Science Inst, RIKEN, Saitama, Japan 351-0198.
$^\ddagger$ Present address: Dept of Physics, Missouri Inst of Science and Technology, Rolla, MO 65409, USA.

\end{document}